# Simulations of the Visible, Infrared and Terahertz Properties of Doped Cyclo[18]carbon Molecules Using Tuned Exchange-Correlation Potential

Jiamin Liu[a], Aigen Li[b], Xianghong Chen[a*], Yonghui Li[a*]

a) Department of Physics and Tianjin Key Laboratory of Low Dimensional Materials Physics and Preparing Technology, School of Science, Tianjin University, Tianjin 300350, China

b) Department of Physics and Astronomy, University of Missouri, Columbia, Missouri 65211, USA

*Corresponding author. E-mail: yonghui.li@tju.edu.cn (Yonghui Li), chenxianghong@tju.edu.cn (Xianghong Chen)

## Abstract

Cyclocarbon molecules are critical in understanding the carbon structure formation and the nature of the interaction between carbon atoms. In cyclocarbons, light elements such as H, O and N may interplay with rings to form doped cyclocarbon molecules. Such molecules show unique optical properties that have never been reported before. In this study, density functional theory with a tuned PBE functional (39% HF exchange) is employed to study the ground and excited states of a cyclo[18] carbon molecule ($C_{18}$) and its doped variants $C_{18}M$ (M = H, Be, B, N, and O). The doping is shown to either make the UV-Vis spectra of $C_{18}$ blue- or red-shifted depending on the spin brought by the dopant. Furthermore, introducing extra-atoms is found to cause the emergence of a new set of infrared modes in the structure under investigation. Finally, applying the molecular dynamic simulations enables one to observe the terahertz characteristics of $C_{18}M$ with frequencies up to 1.5 THz due to the propagation of a particular pattern along the carbon ring.

**Keywords**: cyclo[18]carbon, exchange-correlation potential, UV-Vis spectra, terahertz radiation, molecular dynamics, IR spectra

## I. Introduction

Cyclocarbon molecules, being the third simple carbon substance discovered after spherical carbon molecules [1] and graphene [2], have been devoted to comprehending the structure and properties of pure carbon molecules. Among these, cyclo[18]carbon, or $C_{18}$, was not directly detected via scanning tunneling/atomic force microscopy (STM-AFM) until 2019 [3], although the chemical synthesis of cyclocarbon has begun in 1980s [4]. The traces of $C_{18}$ were also observed in fullerenes [5] and many other structures such as 2D or 3D carbon networks [6]. Furthermore, there are experimental evidences of cyclo[n]carbons [7] in the gas phase [8,9] which may potentially be critical in chemical synthesis.

Over the past years, cyclocarbons have been extensively studied theoretically. Lu *et al.* thoroughly explored the cyclo[18]carbon molecules and reported the alternating bonds in them [10]. According to Stasyuk et al., a cyclo[18]carbon structure possesses a strong electron acceptor characteristic similar to that exhibited by $C_{60}$ [11]. Moreover, theoretical investigations were also applied to $C_{18}$ compounds to elucidate their geometric shape, total energy, electron correlation, aromaticity, tunneling effect, electron transfer, etc [12–18].

Another important question is related to the interactions between the cyclocarbons and other types of molecules. In particular, the synthesis process of a cyclo[18]carbon molecule depends on the detachment of the oxygen atom from the ring [3]. Furthermore, single atomic replacements exert a strong impact on the carbon ring [19], allowing one to adjust their properties due to sensitivity of carbon structures to dopants [20–24]. Thus, a better understanding of doping effects of the cyclocarbon molecules is a relevant task. In fact, the interaction between a single atom and a carbon ring in the gas phase and the associated optical properties are not fully explored. In this respect, the current work particularly focuses on the theoretical investigation of cyclo[18]carbons with extra atoms (light atoms such as H, N and O) attached to the carbon ring. Special attention is paid to ab initio simulations of their electron configurations and optical properties to gain insight into the influence of dopant on the $C_{18}$ structure.

## II. Simulation Method

The geometric structure of a $C_{18}$ molecule has been the subject of debate for a long time. Two possibilities were considered: whether all the bonds in the ring would be of the same length, forming the cumulenic structure, or be alternately short and long, resulting in polyynic chains [4,25]. Such debate is also seen in the theoretical simulations. Density functional theory (DFT) approaches with generalized gradient approximation (GGA) favor the cumulenic structure [26] while the Hartree-Fock (HF) method allows one to obtain the polyynic structure [11]. In that regard, DFT as a "formally exact" route should be tuned for predicting features of excited states and dynamics. Other ways such as multi-configured self-consistent fields [18] and quantum Monte Carlo (QMC) algorithms ensure high-accuracy structures. Therefore, the goal is to calibrate the DFT functionals using the QMC simulation results as benchmarks. According to Torelli et al. [27], the QMC method predicts the polyynic structure of $C_{18}$ with a bond length alternation of 7%.

In DFT approaches, the electron-electron interaction is approximated by a background potential which is the electron density. It comprises the exchange-correlation potential (Vxc) that is the origin of the inaccuracies. The structural differences between PBE and HF suggest the importance of the HF exchange in Vxc. Moreover, numerous theoretical studies of $C_{18}$ and analogous compounds [3,28,29] reveal the importance of the hybrid exchange-correlation functional in the DFT simulation of such structures. Considering the availability and historical performance, we reviewed two variances of PBE functional: PBE0 which contains 25% of exchange and PBE-1/3 which is defined with 1/3 of the HF exchange [30]. The two variances of PBE bring the predicted structures closer to the benchmark but fine-tuning can still be made. In general, the exchange-correlation energy $E_{XC}$ of the hybrid PBE functional can be written as [29]:

$$E_{XC} = \alpha E_X^{Ex} + (1-\alpha) E_X^{PBE} + E_C^{PBE},$$

where $E_X^{PBE}$ and $E_C^{PBE}$ are the exchange and correlations energies and $E_X^{Ex}$ is the HF exchange energy. The parameter α varies between 0 and 1. When α passes a certain threshold, the bond alternation increases significantly. Based on the geometry characters simulated with different values of α shown in **Table 1**, α = 0.39 is the best parameter to match the QMC benchmark.

*Table 1: Structural parameters of cyclo[18]carbon optimized at different percentiles of HF exchange potential in the PBE hybrid functional. Here, r_1 and r_2 are the lengths of adjacent C-C bonds.*

| HF exchange (%) | $r_1$(Å) | $r_2$(Å) | $\bar{r}$(Å) | $\bar{\alpha}$(°) | $\mathcal{A}_r$(%) |
|---|---|---|---|---|---|
| α = 25 | 1.274 | 1.277 | 1.28 | 160 | 0.23 |
| α = 35 | 1.238 | 1.313 | 1.28 | 160 | 5.86 |
| α = 37 | 1.234 | 1.317 | 1.28 | 160 | 6.48 |
| α = 38 | 1.232 | 1.319 | 1.28 | 160 | 6.80 |
| α = 39 | 1.230 | 1.322 | 1.28 | 160 | 7.19 |
| α = 40 | 1.229 | 1.322 | 1.28 | 160 | 7.27 |
| α = 45 | 1.223 | 1.327 | 1.28 | 160 | 8.12 |
| α = 50 | 1.217 | 1.334 | 1.28 | 160 | 9.14 |
| QMC [27] | 1.240 | 1.330 | 1.285 | 160 | 7.00 |

Thus, the 39% PBE hybrid functional was chosen to study the radiation properties of the $C_{18}$ carbon ring. Using this tuned functional, structural relaxations and molecular dynamics (MD) were simulated in Vienna Ab initio Simulation Package (VASP), To avoid interactions between adjacent carbon rings, the simulation box dimensions were 20 Å × 20 Å × 10 Å. The plane waves with a kinetic energy up to 800 eV were employed to expand the electronic wave functions. The atomic structures were considered as fully relaxed when the Hellmann–Feynman forces on each atom were less than 0.02 eV/Å. The UV-Vis and infrared (IR) absorption spectra were simulated using Gaussian 09 software with the same functional (the parameter was added via IOp) and cc-pVTZ basis set was applied to achieve comparable accuracy to VASP.

In this work, attention is mainly paid to the spectral features of doped $C_{18}$ molecules, namely, $C_{18}M$ (M=H, N, O). Their UV-Vis spectra were simulated via the time-dependent DFT method [10,31] and the IR spectra were reproduced using a function of the software based on the DFT-calculated Hessian matrix. Besides, some features were observed within the terahertz range by applying the Fourier transformation to the MD-simulated data. For this, the collisions of the guest atoms with a $C_{18}$ network along the angular direction were imitated by setting the initial velocity at 0.0015 Å/fs and the time step at 2 fs.

# III. Simulation Results

## Energy Surface

**Figure 1**(a) illustrates the optimized structure of pure ($C_{18}$) and doped ($C_{18}M$) molecules (M = H, N, O), simulated with the 39% PBE hybrid functional. In the case of undoped $C_{18}$ molecule, it is seen that the cumulenic structure is predicted to be a transition state between two polyynic structures which can be represented as a saddle point on the energy surface of $C_{18}$ with respect to the two featured bond lengths, $r_1$ and $r_2$ (**Figure 1**(b)). The contour plots show that the two minima correspond to the two equivalent polyynic structures ($r_1 < r_2$ and $r_1 > r_2$). The energy barrier of 0.13 eV, being the difference between the minima and the saddle point, indicates that it is difficult to significantly alter the bond length with low-energy radiations beyond the mid-infrared wavelength range. This result is consistent with data reported by Baryshnikov et al [32]. In turn, the introduction of the extra atom (**Figure 1**(a)) transforms the ring shape into an ellipse, and the C-M bond length gradually decreases with a decrease of the dopant radius in the $C_{18}M$ network. The associated Electron Localization Function (ELF) plots show the transformation of the feature of electrons from polarization to covalent as the number of electrons of the dopant increases. The These structural differences were earlier traced based on the changes in the radii and electronegativity of dopants [19].

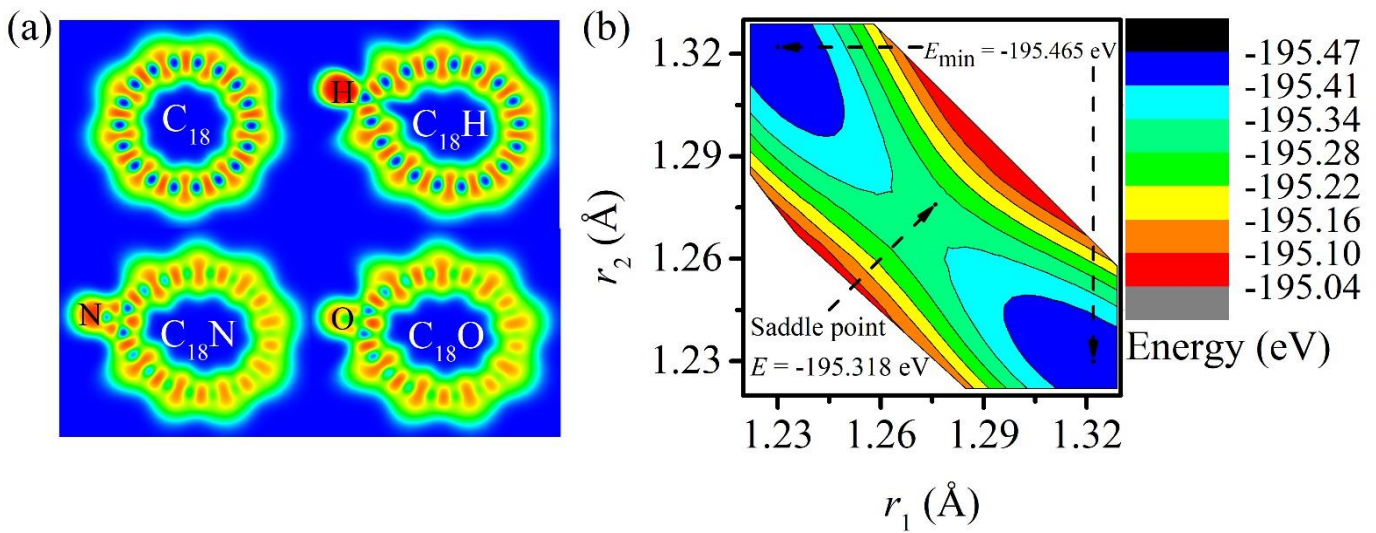

*Figure 1: (a) Optimized structures of $C_{18}$, $C_{18}H$, $C_{18}N$ and $C_{18}O$ molecules with different C-C bond lengths with the associated Electron Localization Function (ELF) represented by color contours. (b) Energy profile of the C18 molecule with respect to the characteristic bond lengths.*

## Capability of the Atom Capture

By selecting two distinct doping sites, the energy profile displays that doping at the edge of the carbon ring achieves superior structural stability. Central adhesions [33] turn to be unstable configurations compare with edge attachments with several eV higher in energy as shown in **Figure 2**. On the side attachments, one can consider the side adhesion process as a chemical reaction featured by intermediate states (stables states in **Figure 2**) and transition states (transition states in **Figure 2**). Each single atom must overcome energy barriers before it can stay to a region with minimum energy. The unoccupied p-orbital of N atoms supports the formation of σ-bonds with the C18 carbon ring, which contributes to a reduction in the system's total energy. The pronounced electronegativity of the O atom induces notable localized strain, leading to a decrease in molecular energy with a significant energy barrier. The covalent bond between the H atom and carbon introduces only minor perturbations to the carbon ring structure, maintaining a relatively steady impact on energy. Conversely, due to Be minimal electronegativity, it results in the highest stable state energy.

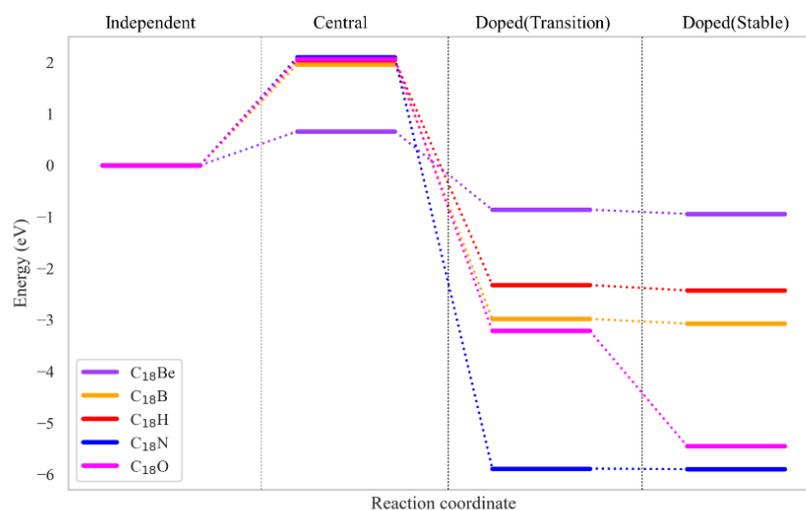

*Figure 2: The comparison of energy when doping atom M is placed at the central site of the carbon ring versus the edge site, as well as the energy variation of intermediate states during the more stable doping process at the edge site, is illustrated. "Central" refers to the central site of the carbon ring, while "Doped" signifies the edge site.*

The Fukui function affords a rigorous quantitative approach to pinpoint areas within a molecule that are most susceptible to chemical reactions. Observations from **Figure 3** (left panel) elucidate a unique 3D electron cloud distribution when the carbon ring is doped with a Be atom. Based on the Fukui charge definition, it's evident that upon electron supplementation to the Be-doped carbon ring, it manifests pronounced reactive zones throughout, more conspicuous than rings doped with other elements, resulting in the highest energy at its edge-doped sites (**Figure 2**). The rightward electron migration in the Be-doped carbon ring can be ascribed to the repulsion between the newly introduced electron and the extant electrons, compounded by the electron cloud redistribution owing to the σ-bond established with the Be atom. The heightened electronegativity of the O atom renders it adept at electron sequestration; the vacant p-orbital of the N atom culminates in a σ-bond with the carbon ring. The influence of the H atom on the carbon ring is minimal, attributable to its atomic radius and spatial configuration. These collective dynamics result in the N, H, B, and O atoms manifesting distinct reactivity domains within their electron density distribution. Moreover, when the composite system loses an electron (**Figure 3**, right panel), the reactivity zones for H and Be predominantly concentrate on the end distal to the doped atom, whereas for B, N, and O, the zones are proximal to the doped atom.

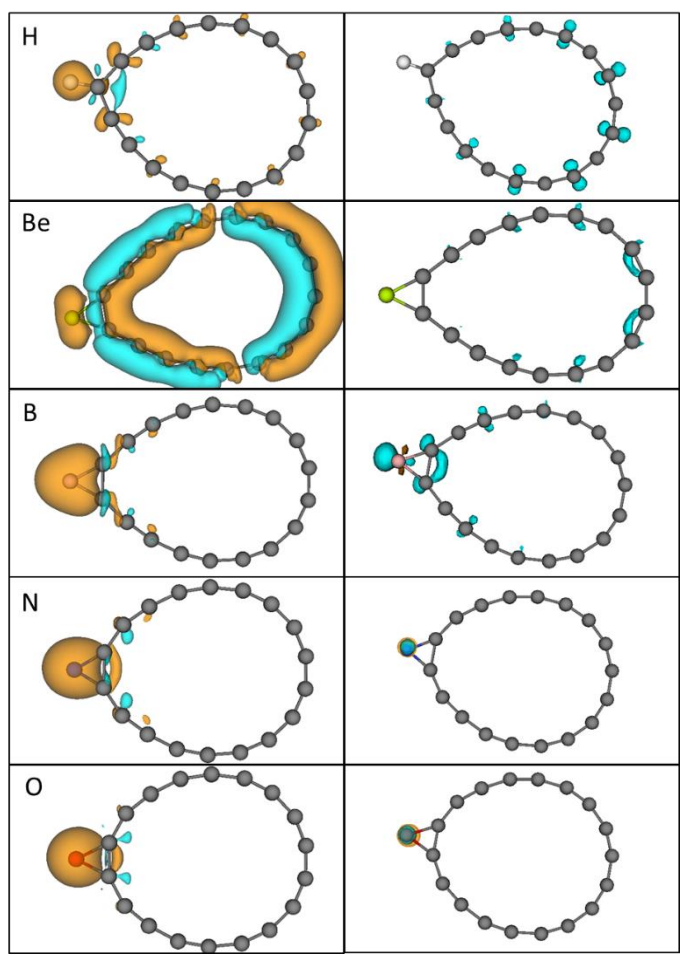

*Figure 3: Fukui charge, Intuitive information on how the five doping atoms affect the structure and electron cloud distribution of the $C_{18}$ carbon ring. The orange and blue electron clouds depict the increase and decrease in electron density, respectively. The activation energy and the geometry in the transition states can be used to describe the potential of adhesion of the extra atoms.*

**Figure 4** displays the simulated UV-Vis absorption spectra of the molecules under consideration. For pure $C_{18}$, it exhibits an extremely strong absorption peak at 231.2 nm. In turn, the M-atom-doped rings reveal the drastic spectral shifts. If the dopant possesses an unpaired spin, the corresponding spectra are red-shifted, as seen in the case of $C_{18}H$, $C_{18}B$, and $C_{18}N$, whose absorption peaks move towards 240 nm, 270 nm, and 272 nm, respectively. In contrast, the UV-Vis spectra of $C_{18}Be$ and $C_{18}O$ are blue-shifted since these molecules still have a closed-shell electronic structure. Furthermore, all $C_{18}M$ molecules also exhibit weak long-wavelength peaks, which result from the reduction of the optical gap due to the extra energy level of the guest atoms. More details of the spectra are provided in Table S1 and S2.

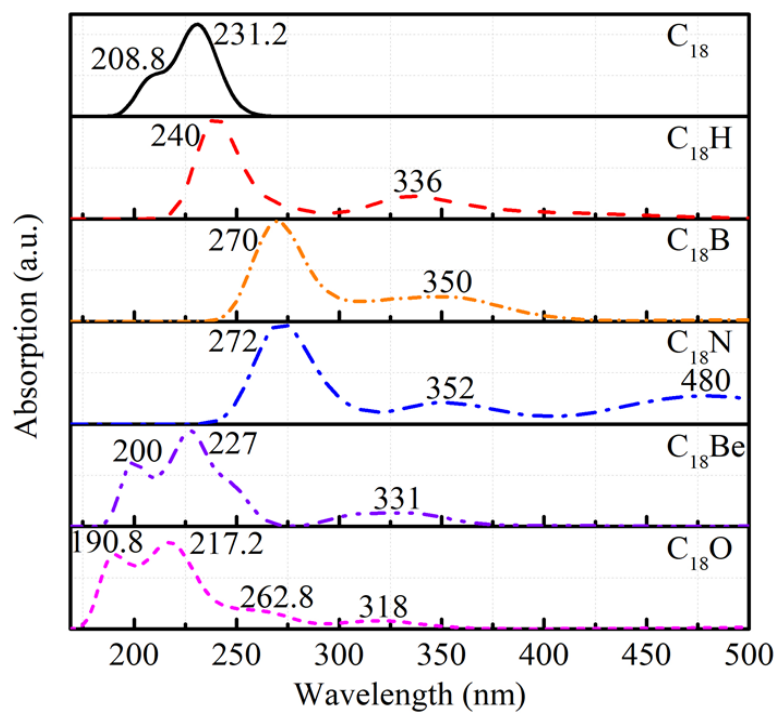

**Figure 4:** Simulated UV-Vis absorption spectra of $C_{18}$ and $C_{18}M$ molecules

Besides the electronic oscillation data provided by the UV-Vis spectra, the molecular vibrations were also analyzed by simulating the IR spectra of $C_{18}$-based structures. As shown in **Figure 5**, the first peak of a $C_{18}$ molecule, arising at 2164 cm$^{-1}$, is assigned to a C-C stretching vibration (here and hereinafter referred to as Mode I). The second peak at 496 cm$^{-1}$ corresponds to the C-C in-plane bending (here and hereinafter referred to as Mode II). The results are consistent with those reported by Liu et al [34] but the peak intensities in this work were about 5% higher.

With the introduction of H, N and O atoms, the $C_{18}M$ structure not only retains the two original classes of C-C vibration modes but also exhibits the emergence of new bands (here and hereinafter referred to as Mode III). For $C_{18}H$, the Mode I and II oscillations are red-shifted to 1744 and 480 cm$^{-1}$, respectively. The Mode III features associated with C-H vibrations emerge at 1012 and 598 cm$^{-1}$ respectively. Unlike the $C_{18}H$ network, the N atom doping made the Mode I split into two modes at 2182 and 2116 cm$^{-1}$. In the IR spectrum of $C_{18}N$, the Mode II peak is red-shifted to 352 cm$^{-1}$. Besides, the Mode III oscillations of $C_{18}N$, associated with the motion of the N atom, are located at 1870, 1582, and 898 cm$^{-1}$, respectively. The intercalation of the O atom leads to a red-shift of the Mode I to 2122 cm$^{-1}$ and the splitting of the Mode II into two peaks at 514 and 376 cm$^{-1}$. The strongest Mode III peak of $C_{18}O$ arises at 922 cm$^{-1}$ and is followed by a signature at 988 cm$^{-1}$. Therefore, the dopants significantly change the oscillation patterns of the original cyclocarbon structure, causing the red-shift of the IR peaks along with a peak splitting. The guest atom generally creates pronounced Mode III oscillations.

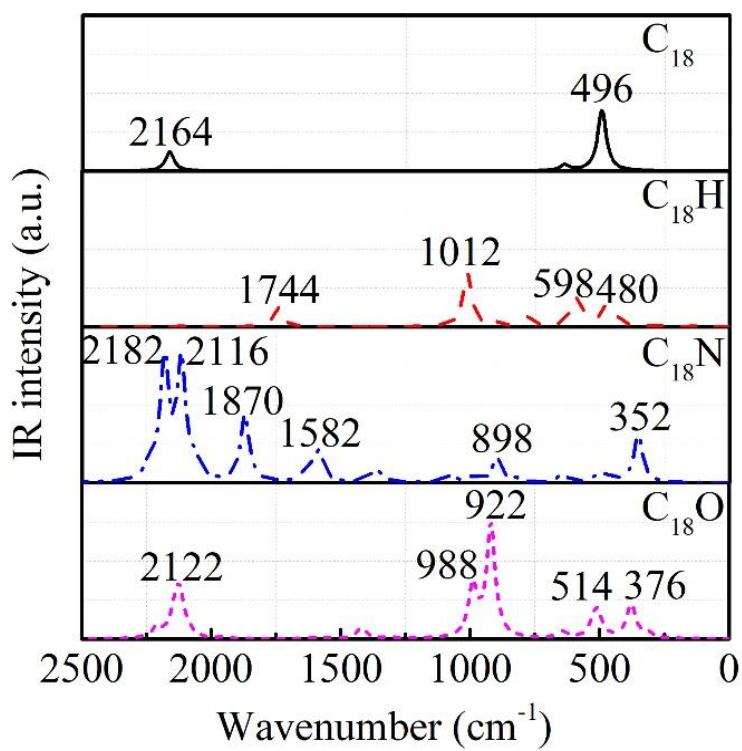

*Figure 5: Simulated IR spectra of $C_{18}$ and $C_{18}M$ molecules*

The spectra during the formation of the $C_{18}M$ molecules were also simulated via MD algorithms. The collision of the extra atom (H, N, or O) with a carbon ring results in a particular movement pattern. This can be extracted using the distance between the guest atom and the center of mass of the complex, appearing as the periodical oscillation in **Figure 6**(a) for each case. Similar behaviors are also observed on each carbon atom in the ring. The overall structure of the $C_{18}$ molecule exhibits the regular fluctuations with a period of approximately 800 fs which corresponds to the terahertz radiation. In turn, **Figure 6**(b) displays the corresponding Fourier transform spectra of the patterns from **Figure 6**(a), where the characteristic peaks are clearly seen around 1.5 THz, being beyond the IR modes. (Such character of motion is seen on other atoms as shown in Figure S1).

However, the terahertz peaks released by the carbon ring slightly shift with doping by H, N, and O atoms. This is due to the propagation of the motion pattern along the carbon ring which does not significantly depend on the guest atom.

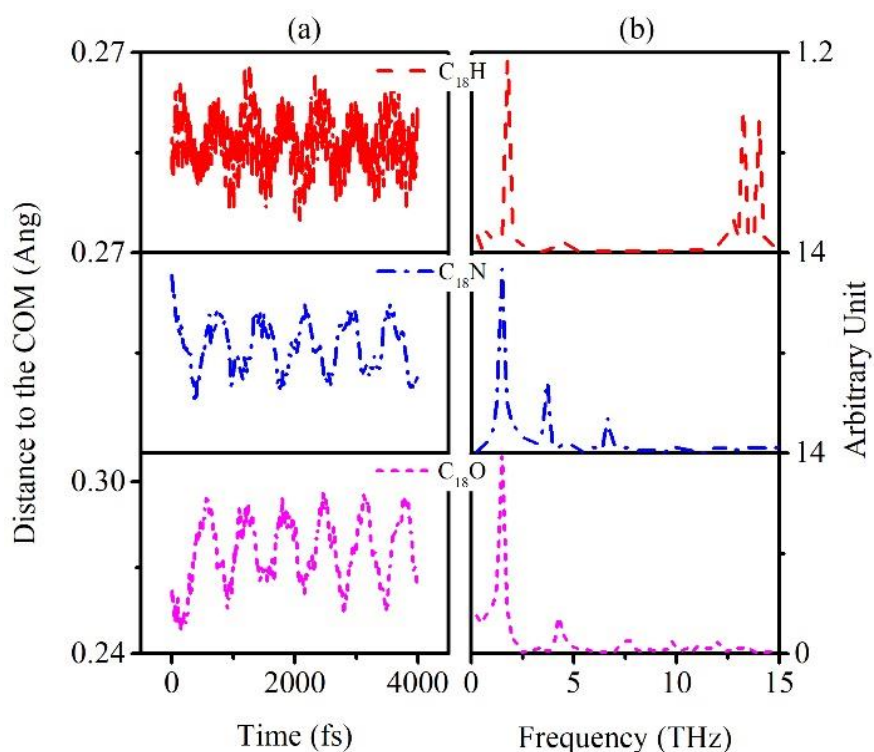

***Figure 6:*** *(a) Distance between M (M = H, N, or O) atoms and the center of mass evolving with time. (b) Fourier transform spectra of patterns from Figure (a).*

## IV. Summary

In this work, the tuned 39% PBE hybrid functional was introduced to simulate the cyclocarbon molecule $C_{18}$ and its doped variances, $C_{18}M$ (M = H, Be, B, N, and O). According to out simulations, the introduction of M atoms transforms the original ring structure into the ellipse structures. Furthermore, the C-M bond extends as the dopant radius increase. Special attention was paid to the optical properties of the compounds under consideration, namely, the impacts of the dopants on the UV-Vis, IR and terahertz spectra. First, the UV-Vis spectra were shown to depend on the extra-atom spin. In particular, embedding the atoms with unpaired spins caused the red-shifting of the spectra. While the UV-Vis spectra were blue-shifted if atoms with paired spin were doped. Second, the IR spectra of the doped cyclo[18]carbons comprised the two sets of vibrational modes inherited from the original carbon ring and one new set of modes induced by the interaction between the doped atom and the ring. Finally, the radiation features arising in the collisions between the cyclocarbon ring and the guest atoms during the formation of a $C_{18}M$ structure were simulated via MD algorithms. The impacts of the H, N, and O atoms on the cyclo[18]carbon rings were shown to cause a stable 1.5 THz peak due to the propagation of the motion pattern along the ring.

## Declarations

**Funding**
Yonghui Li was sponsored by the National Key Research and Development Program of China (Grant No.2021YFF1200701), the National Natural Science Foundation of China (Grant No.11804248) and the Key Projects of Tianjin Natural Fund 21JCZDJC00490.

**Conflicts of interest**
The authors confirm that there are no known conflicts of interest associated with this publication and there has been no significant financial support for this work that could have influenced its outcome.

**Data Availability Statement**

Data available in article or supplementary material.